\documentclass[%
nofootinbib,
 amsmath,amssymb,
 aps,
twocolumn
]{revtex4}

\usepackage{graphicx}
\usepackage{dcolumn}
\usepackage{bm}
\usepackage[unicode,hypertexnames,setpagesize,%
pdftex,%
colorlinks,%
citecolor=blue,%
hyperindex,%
plainpages=false,%
bookmarksopen,%
bookmarksnumbered%
]{hyperref}

\usepackage{color}

\newcommand{\colvec}[2]{%
	\scalebox{0.8}{%
		\renewcommand{\arraystretch}{.8}%
		$\begin{Bmatrix}#1 \\ #2  \end{Bmatrix}$%
	}
}
\global\long\def\christoffel#1#2{\colvec{#1}{#2}}%

\begin{document}

\preprint{APS/123-QED}

\title{LyST: a Scalar-Tensor Theory of Gravity on Lyra Manifold }
\thanks{Dedicated to Prof. Ruben Aldrovandi}%

\author{R. R. Cuzinatto}
\affiliation{%
 Instituto de Ci\^{e}ncia e Tecnologia, Universidade Federal de Alfenas, Rodovia Jos\'{e} Aur\'{e}lio Vilela, 11999, Po\c{c}os de Caldas, MG, Brazil
}%

\author{E. M. de Morais}
\author{B. M. Pimentel}%
\affiliation{%
	Institute of Theoretical Physics, S\~{a}o Paulo State University, 
	R. Dr. Bento Teobaldo Ferraz 271, S\~{a}o Paulo, SP, Brazil
}%

\date{\today}

\begin{abstract}
We present a scalar-tensor theory of gravity on a torsion-free and
metric compatible Lyra manifold. This is obtained by generalizing
the concept of physical reference frame by considering a scale function
defined over the manifold. The choice of a specific frame induces
a local base, naturally non-holonomic, whose structure constants give
rise to extra terms in the expression of the connection coefficients
and in the expression for the covariant derivative. In the Lyra manifold,
transformations between reference frames involving both coordinates
and scale change the transformation law of tensor fields, when compared
to those of the Riemann manifold. From a direct generalization of
the Einstein-Hilbert minimal action coupled with a matter term, it
was possible to build a Lyra invariant action, which gives rise to
the associated Lyra Scalar-Tensor theory of gravity (LyST), with field
equations for $g_{\mu\nu}$ and $\phi$. These equations have a well-defined
Newtonian limit, from which it can be seen that both metric and scale
play a role in the description gravitational interaction. We present a spherically symmetric solution for the LyST gravity field equations. It dependent on two parameters $m$ and $r_{L}$, whose physical meaning is carefully investigated. We highlight the properties of LyST spherically symmetric line element and compare it to Schwarzchild solution.

\end{abstract}

\maketitle


\section{Introduction}

In the formulation of General Relativity, Einstein presented a geometric
structure resulting from a set of adopted hypotheses, among which
are the adoption of Riemannian manifold for modeling space-time, the
shape of space-time infinitesimal interval, the integrability of vector
lengths and the absence of torsion \cite{einstein1916foundation}.
This fact opens avenues for formulating alternative theories of gravitaty
through changing these considerations \cite{kaluza1921unitatsproblem,klein1926quantentheorie,finsler1918,einstein:1928:RGA}.
In this scenario, Hermann Weyl's unified theory, published in $1918$,
gains prominence, since it raises the possibility of using the freedom
to adopt these geometric hypotheses to formulate a structure that
accommodates both gravitation and electromagnetism \cite{weyl1918gravitation}.
In Weyl geometry, the change in the length of vectors under parallel
transport is non-zero and depends on a new vector quantity $A_{\mu}$
that plays the role of electromagnetic potential. An important finding
is that the presented structure exhibits a new type of symmetry, called
gauge symmetry, in addition to invariance by diffeomorphisms.

In spite of its simplicity, mathematical beauty and great potential
for unifying fundamental theories based on simple geometric concepts,
Weyl's theory presents problems of a physical nature. In the first
place, as pointed out by Einstein, the hypothesis of non-integra-bility
of length makes the frequency of spectral lines emitted by atoms depend
on their past history and, as such, would not remain constant \cite{goenner2004history}.
Moreover, the Lagrangian density invariance under diffeomorphisms
and gauge transformations gives rise to fourth order field equations,
which is not desirable in a physical theory \cite{pagani1987problem}.
Notwithstanding, Weyl's work is widely recognized, since he was the
pioneer in the approach to gauge theories, a concept on which much
of the work in modern physics is based. One way to reduce the problems
inherent in Weyl's theory is to impose that the Weyl displacement
vector is irrotational, that is, $A_{\mu}=\partial_{\mu}\sigma$ \cite{weyl1922space}.
Theories with this characteristic, called WIST (Weyl Integrable SpaceTime)
have attracted the attention of researchers in recent years  \cite{scholz2018unexpected}.
In this approach, the unification between gravity and electromagnetism
is set aside, and WIST is seen as a scalar-tensor theory of gravity.

Another way of maintaining the vector length integrability was proposed
by Lyra in 1951 through the adoption of a gauge function $\phi$ as
an intrinsic part of the manifold's geometric structure  \cite{lyra1951modifikation}.
This approach naturally alters the definition of physical frames,
since these, in Lyra's geometry, depend on both coordinates and the
gauge function. Lyra proposes that the affine connection describing
parallel transport be defined by the sum of two sectors; one of them
depending on the metric, and the second one depending exclusively
on a quantity $\phi$, such that the curvature tensor, the torsion
tensor and their respective contracted forms will be functions, not
only of the metric, but also of $\phi$. 

Although he described the geometric structure, Lyra did not formulate
a field theory where geometric objects play the role of gravitational
field. The first proposal to address the matter was made by D. K.
Sen, in 1957, in the formulation of a static cosmological model where
the scale function appears as responsible for the redshift of the
galactic spectral lines \cite{sen1957static}. For that, Sen proposed
to obtain the field equations through the variational principle: 
\begin{equation}
\delta\left(\int d^{4}x\phi^{4}\sqrt{-g}\mathcal{R}\right)=0\,,\label{eq:000}
\end{equation}
where both the integration element $d^{4}x\phi^{4}$ and the scalar
curvature $\mathcal{R}$ are invariant under Lyra's reference frame
transformation. The equations presented as a result were:
\begin{equation}
\mathcal{R}_{\mu\nu}-\frac{1}{2}g_{\mu\nu}\mathcal{R}+\frac{3}{2}A_{\mu}A_{\nu}-\frac{3}{4}g_{\mu\nu}A_{\lambda}A^{\lambda}=\varkappa T_{\mu\nu}\label{eq:00}
\end{equation}
where $A_{\mu}$ is a vector field that, according to the author,
is a direct consequence of the gauge function $\phi$ parallel transport,
although he does not make it clear how $A_{\mu}$ is related to $\phi$.
In this equation, $\mathcal{R}_{\mu\nu}$ and $\mathcal{R}$ are the
Ricci tensor and the scalar curvature calculated with Christoffel
symbols, as in General Relativity. In order to derive  (\ref{eq:00}),
Sen used the gauge fixing condition $\phi=1$, keeping the components
of $A_{\mu}$ free and varying the action with respect to the metric
components. Subsequently, Sen showed that geodesics on Lyra manifold
are not, in general, autoparallel curves \cite{sen1960geodesics}.
However, as a specific case, it is possible to guarantee the autoparallelism
of the geodesics by imposing the condition $A_{\mu}=\phi^{-1}\partial_{\mu}\ln\phi^{2}$.
After the measuring of the cosmic background radiation and confirmation
of the Big-Bang model \cite{penzias1965measurement}, Eq. (\ref{eq:00})
was widely used for modeling the cosmological dynamics \cite{bhamra1974cosmological,beesham1986friedmann,halford1970cosmological,reddy1988magnetized,singh1997new,singh2008exact}.

Later, in $1971$, Sen and Dunn took a step further in an attempt
to formalize a gravity theory in Lyra manifold \cite{sen1971scalar}.
To this end, the authors, recognized the simultaneous influence of
$g_{\mu\nu}$ and $A_{\mu}$ on gravitational phenomena and performed
the variational procedure without fixing the gauge. Instead, they
varied the action with respect to the metric components and the components
of $A_{\mu}$. The field equation from metric variations recovers
(\ref{eq:00}) in the limit $\phi\rightarrow1$. The second set of
equations, given by 
\begin{equation}
3A_{\mu}+\frac{3}{2}\phi^{-1}\partial_{\mu}\ln\phi^{2}=0\,,\label{eq:00c}
\end{equation}
directly relates $A_{\mu}$ to $\phi$. This equation is problematic.
First, it is explicitly incompatible with the condition of geodesic
autoparallelism, and, as a consequence, metric geodesics and affine
geodesics do not coincide in the gravity theory formulated by Sen
\& Dunn. Secondly, the gauge-fixing condition used in the derivation
of (\ref{eq:00}), namely $\phi=1$, leads to the vanishing of
$A_{\mu}$. In other words, by imposing the gauge condition in the
Sen \& Dunn analogue of Einstein's equations, one recovers (\ref{eq:00}).
However, the application of the same condition on (\ref{eq:00c})
causes $A_{\mu}=0$. So $\phi=1$ does not act as a gauge-fixing condition,
but most properly as a General Relativity limit condition for Sen
\& Dunn gravity. In fact, disregarding the gauge-fixing condition
in this theory makes it possible to work with coupled differential
equations for the metric coefficients and the $\phi$ function. In
this case, Sen \& Dunn set aside Eq. (\ref{eq:00}) and combined the
two sets of equations in a single expression:
\begin{equation}
\mathcal{R}_{\mu\nu}-\frac{1}{2}g_{\mu\nu}\mathcal{R}-\frac{3}{2\phi^{2}}\phi_{,\mu}\phi_{,\nu}+\frac{3g_{\mu\nu}}{4\phi^{2}}\phi_{,\lambda}\phi^{,\lambda}=\varkappa\phi^{2}T_{\mu\nu}\,.\label{eq:00b}
\end{equation}
This is an interesting expression, since it can be interpreted as
a specific case of Brans-Dicke's theory \cite{brans1961mach}. The
authors proposed a spherically symmetric solution for (\ref{eq:00b})
written in terms of power series \cite{sen1971scalar}. The following
year, Halford found a spherically symmetric analytical solution in
isotropic coordinates \cite{halford1972scalar}. 

In 1975, Jeavons, McIntosh and Sen pointed out the heuristic importance
of Eq. (4) but showed that it can not be obtained from the principle
of least action \cite{jeavons1975correction}, unlike what is proposed
in \cite{sen1971scalar}. This equation was obtained by neglecting
the contribution of the terms $\int d^{4}x\phi^{2}\delta\left(\sqrt{-g}\mathcal{R}\right)$
and $\int d^{4}x\phi^{3}\delta\left(\left(\sqrt{-g}A_{\mu}g^{\mu\nu}\right)_{,\nu}\right)$.
The field equation that respects the aforementioned principle, arises
from the variation of the action with respect to the components of
the metric tensor, while the condition of self-parallelism of geodesics
$A_{\mu}=\phi^{-1}\partial_{\mu}\ln\phi^{2}$ arises naturally from
the variation of the action with respect to $A_{\mu}$. 

Through the knowledge on theories of gravity in the Lyra manifold,
one can notice that the presence of a scale in the vector transformation
law naturally induces extra terms to Christoffel symbols. However,
the introduction of a vector field $A_{\mu}$ as an entity uncorrelated
to $\phi$ describing these terms does not seem a strictly necessary
procedure, since the field equations themselves establish a direct
relation between $A_{\mu}$ and $\phi$, both in the work of Sen \&
Dunn of $1971$ and in the theory of Jeavons, McIntosh \& Sen of 1975.
Thus, this work proposes a new interpretation on the description of
the Lyra manifold, assuming $\phi$, and the metric tensor as the
fundamental fields of a Scalar-Tensor Theory of Gravity, here called
LyST, a shorthand notation for Lyra Scalar-Tensor Theory of Gravity.
In order to do so, in Section \ref{sec:Geometria-de-Lyra} we build
a manifold where the scale function $\phi$ and the coordinates define
a Lyra reference. The choice of a specific reference frame induces
a local basis, naturally non-holonomic, where the structure constants
of the Lie algebra depend exclusively on $\phi$. This geometric structure
leads naturally to Lyra's vector transformation law and induces extra
terms in the related affine connection. Once this framework is established,
the Lyra manifold is adopted as a space-time model, in Section \ref{sec:Espa=0000E7o-Tempo-de-Lyra},
where we impose suitable geometric conditions and calculate curvature
properties. In the Section \ref{sec:Equacoes-de-Campo}, a variational
principle is proposed through which field equations can be obtained
for the metric coefficients and the scale function. Such equations
have an appropriate Newtonian limit, as shown in Section \ref{subsec:O-Limite-Newtoniano}. A symmetrically symmetrical solution for LyST gravity is presented in Section \ref{sec:vac-sph}. The vacuum solution is built up in Section \ref{sec:5.1}. It is shown in Section \ref{sec:Propriedades-da-solucao} that it reduces to the ordinary Schwarzschild line element in the appropriate limit. The long-distance regime of LyST spherically symmetric solution accommodates the tantalizing possibility of reproducing a cosmological costant-like behaviour, either in the ways of a de-Sitter solution or similar to a Anti-de-Sitter space-time. We identify two different classes of LyST spherically symmetric metrics in Section \ref{sec:5.3}. Section \ref{sec:Singularidades-e-Estrutura} is dedicate to study the singularities of this metric and to discuss its causal structure. Section \ref{sec:final} contains our final comments and future perspectives.

\section{Lyra Geometry \label{sec:Geometria-de-Lyra}}

A Lyra differential manifold $\mathcal{M}$ is a second-countable
Hausdorff topological space of dimension $n$. A reference frame in
this geometry is represented by the triad $\left(U_{i},\chi_{i},\Phi_{i}\right)$,
where $U_{i}\subset\mathcal{M}$ is a open subset of $\mathcal{M}$,
$\left\{ \left.\chi_{i}:U_{i}\to\mathbb{R}^{n}\right|n\in\mathbb{Z}_{+}\right\} $
is a map parameterizing the coordinates and $\Phi:U_{i}\to\mathbb{R}^{*}$
a scale map. Thus, considering $P\in\mathcal{M}$, then the coordinates
of $P$ are defined as the map $x:=\chi\circ P:\mathcal{M}\to\mathbb{R}^{n}$
and the scale function as $\phi:=\Phi\circ\chi^{-1}:\mathbb{R}^{n}\to\mathbb{R}$.
Therefore, $\Phi\circ P=\Phi\circ\chi^{-1}\circ\chi\circ P$ can be
written as $\phi\left(x_{P}\right):=\phi\circ x_{P}$.

The reference frames $\left\{ \left(U_{i},\chi_{i},\Phi_{i}\right)\right\} $
on $\mathcal{M}$ must respect \cite{kobayashi1963foundations}:

\textbf{(i)} $\bigcup_{i\in I}U_{i}=\mathcal{M}$, where $i$ is an
index that covers a total account $I$ of frames~;

\textbf{(ii)} the maps $\chi_{i}\circ\chi_{j}^{-1}$ and $\chi_{j}\circ\chi_{i}^{-1}$
must be $C^{\infty}$ functions in their domains whenever $U_{i}\cap U_{j}\neq  \emptyset$ \footnote{The set $C^{\infty}$ consists of all differentiable functions $f:\mathcal{M}\to U$ in $P$.};

\textbf{(iii)} the maps $\Phi_{i}\circ\chi_{i}$ must be $C^{\infty}$
functions in $U_{i}$ for each $i\in I$; and

\textbf{(iv)} the family of reference frames is maximal with respect
to \textbf{(i)}, \textbf{(ii)} and \textbf{(iii)}.

\noindent 

Consider a curve $\alpha\left(t\right)$ over $\mathcal{M}$ through
the differential map $\alpha:\mathbb{R}\to\mathcal{M}$, where $\alpha\left(0\right):=P\in\mathcal{M}$,
and a $C^{\infty}$ function $f:M\to\mathbb{R}$. It is possible to
define a \emph{tangent vector} $\mathbf{v}_{\left(\alpha\right)}$
in $P$, as a map $\mathbf{v}_{\left(\alpha\right)}:C^{\infty}\to\mathbb{R}$
expressed by:
\begin{equation}
\mathbf{v}_{\left(\alpha\right)}f=\left.\frac{d\left(f\circ\alpha\right)}{dt}\right|_{P}\,.\label{eq:01}
\end{equation}
By considering an specific reference frame $\left(U,\chi,\Phi\right)$,
Eq. (\ref{eq:01}) can be represented by $\mathbf{v}_{\left(\alpha\right)}f=x^{\mu\prime}\left(0\right)\partial_{\mu}f$.

In Lyra's geometry, the coordinates $x^{\mu}$ and scale function
$\phi$ related to the choice of a specific reference frame can be
used to define a natural basis as:
\begin{equation}
\mathbf{e}_{\mu}=\frac{1}{\phi\left(x\right)}\boldsymbol{\partial}_{\mu}.\label{eq:02}
\end{equation}
Given the basis, any tangent vector $\mathbf{v}_{\left(\alpha\right)}$
can be represented as $\mathbf{v}_{\left(\alpha\right)}=v_{\left(\alpha\right)}^{\hphantom{\alpha}\mu}\mathbf{e}_{\mu}$,
where $v_{\left(\alpha\right)}^{\hphantom{\alpha}\mu}$, given by:
\begin{equation}
v_{\left(\alpha\right)}^{\hphantom{\alpha}\mu}:=\phi\left(x\right)x^{\mu\prime}\left(0\right)\label{eq:03}
\end{equation}
are the components of $\mathbf{v}_{\left(\alpha\right)}$. The set
$\left\{ \mathbf{v}_{1},\cdots,\mathbf{v}_{K}\right\} $ of all tangent
vectors in $P$ forms a tangent vector space $\mathbf{T}_{P}\mathcal{M}=\bigcup_{k\in K}\mathbf{v}_{k}$.\emph{ }

A direct consequence of the expression (\ref{eq:02}) is the non-commutativity
of the basis elements induced by the choice of a reference frame.
Indeed, the Lie algebra can be characterized by:

\begin{equation}
\left[\mathbf{e}_{\mu},\mathbf{e}_{\nu}\right]=\phi^{-2}\left(\delta_{\mu}^{\alpha}\partial_{\nu}\phi-\delta_{\nu}^{\alpha}\partial_{\mu}\phi\right)\mathbf{e}_{\alpha}.\label{eq:03b}
\end{equation}
The fact that the structure constants are non-null leads to important
consequences to the geometric properties of the manifold. It is convenient
to define the notation $\gamma_{\hphantom{\alpha}\mu\nu}^{\alpha}=\phi^{-2}\left(\delta_{\mu}^{\alpha}\partial_{\nu}\phi-\delta_{\nu}^{\alpha}\partial_{\mu}\phi\right)$.

Let $\left(U,\chi,\Phi\right)$ e $\left(\bar{U},\bar{\chi},\bar{\Phi}\right)$
two reference frames, such that $U\cap\bar{U}\neq \emptyset $
and $P\in U\cap\bar{U}$. The relation between $\mathbf{e}_{\mu}$
e $\bar{\mathbf{e}}_{\mu}$, according to (\ref{eq:02}), is:
\begin{equation}
\bar{\mathbf{e}}_{\mu}=\frac{\phi\left(x\right)}{\bar{\phi}\left(\bar{x}\right)}\frac{\partial x^{\nu}}{\partial\bar{x}^{\mu}}\mathbf{e}_{\nu}.\label{eq:04}
\end{equation}

The components of the tangent vector $\mathbf{v}_{\left(\alpha\right)}\in\mathbf{T}_{P}\mathcal{M}$
in reference frame $U$ relate to its components in reference $\bar{U}$
via \cite{lyra1951modifikation}:

\begin{equation}
\bar{v}_{\left(\alpha\right)}^{\hphantom{\alpha}\mu}=\frac{\bar{\phi}\left(\bar{x}\right)}{\phi\left(x\right)}\frac{\partial\bar{x}^{\mu}}{\partial x^{\nu}}v_{\left(\alpha\right)}^{\hphantom{\alpha}\nu}\,.\label{eq:05}
\end{equation}

A covector $\boldsymbol{\omega}$ over the point $P\in\mathcal{M}$
is defined as a linear map $\boldsymbol{\mathbf{\omega}}:\mathbf{T}_{P}\mathcal{M}\to\mathbb{R}$,
which takes tangent vectors as arguments. The set of all covectors
in $P$ form a vector space, called a cotangent space $\mathbf{T}_{P}^{*}\mathcal{M}$.
A covector $\boldsymbol{\omega}$ can be represented in a specific
basis, denoted by $\left\{ \boldsymbol{\theta}^{\nu}\right\} $, as
$\boldsymbol{\omega}=\omega_{\nu}\boldsymbol{\theta}^{\nu}$. Applying
this in a given vector $\mathbf{v}\in\mathbf{T}_{P}\mathcal{M}$,
the result is given by: 
\begin{equation}
\boldsymbol{\omega}_{P}\circ\mathbf{v}_{P}=\left(\omega_{\nu}\boldsymbol{\theta}^{\nu}\right)\circ\left(v^{\mu}\mathbf{e}_{\mu}\right)=v^{\mu}\omega_{\mu}\label{eq:06}
\end{equation}
where $\omega_{\mu}:=\boldsymbol{\omega}\circ\mathbf{e}_{\mu}$ are
the components of the covector $\boldsymbol{\omega}$ and it was assumed
the orthonormality condition $\mathbf{\boldsymbol{\theta}}^{\nu}\circ\mathbf{e}_{\mu}=\delta_{\mu}^{\nu}$
to the basis. 

Let a curve $\alpha\left(t\right)$ in $\mathcal{M}$ given by the
differentiable map $\alpha:\mathbb{R}\to\mathcal{M}$, with $\alpha\left(0\right):=P\in\mathcal{M}$
and a smooth function $f:\mathcal{M}\to\mathbb{R}$. Thus, one can
check that the map $f\circ\alpha:\mathbb{R}\to\mathbb{R}$, where
$f\circ\alpha=f\circ\chi^{-1}\circ\left(\chi\circ\alpha\right)=\tilde{f}\left(x\left(t\right)\right)$
is an ordinary function of $t$ that describes the behaviour of $f$
along the curve. Taking the derivative of this function in P, cf.
(\ref{eq:01}), it turns out that the value of $f$ at t = 0 depends
only on the tangent vector $\mathbf{v}_{\left(\alpha\right)}$ to
the $\alpha$ curve. That is, the operation $\left.d\left(f\circ\alpha\right)/dt\right|_{P}=\mathbf{v}_{\left(\alpha\right)}f$
defines a vector $\mathbf{d}f:\mathbf{T}_{P}\mathcal{M}\to\mathbb{R}$.
By choosing a reference frame, $\mathbf{d}f$ is represented by:
\begin{equation}
\mathbf{d}f\circ\mathbf{\mathbf{v}}=v^{\lambda}\phi^{-1}\partial_{\lambda}\tilde{f},\label{eq:06a}
\end{equation}
whence, according to the Eq. (\ref{eq:06}), $\phi^{-1}\partial_{\mu}f$
are the components of $\mathbf{d}f$.

As mentioned above, a basis $\boldsymbol{\theta}^{\mu}$ to cotangent
space $\mathbf{T}_{P}^{*}\mathcal{M}$ is obtained by requiring the
orthonormality condition. This is done by defining $n$ smooth functions
$x^{\mu}:\mathcal{M}\to\mathbb{R}$, parameterized in a specific reference
frame which receive a point in $\mathcal{M}$ as argument and return
the $\mu-$th coordinate. Let us take $\tilde{f}$ as $x^{\mu}$ and
$\mathbf{v}$ as $\mathbf{e}_{\nu}$ in Eq. (\ref{eq:06a}). It follows
that $\mathbf{d}x^{\mu}\circ\mathbf{e}_{\nu}=\delta_{\mu}^{\lambda}\phi^{-1}\partial_{\lambda}x^{\nu}=\delta_{\mu}^{\nu}\phi^{-1}$.
Consequently, an adequate basis for $\mathbf{T}_{P}^{*}\mathcal{M}$
will be:
\begin{equation}
\boldsymbol{\theta}^{\mu}=\phi\mathbf{d}x^{\mu}.\label{eq:07-1}
\end{equation}
So, a general covector $\boldsymbol{\omega}$ can be written as $\boldsymbol{\omega}=\omega_{\mu}\phi\mathbf{d}x^{\mu}$
in a local frame. Under a change of frames, its components are transformed
as:
\begin{equation}
\bar{\omega}_{\mu}=\frac{\phi\left(x\right)}{\bar{\phi}\left(\bar{x}\right)}\frac{\partial x^{\nu}}{\partial\bar{x}^{\mu}}\omega_{\nu}.\label{eq:08-1}
\end{equation}

The tensors $\mathbf{W}$ of type $\left(p,q\right)$ are defined
as applications receiving $p$ tangent covectors and $q$ tangent
vectors, mapping them to a real number: 
\[
\mathbf{W}:\underbrace{\mathbf{T}_{P}^{*}\mathcal{M}\mathcal{\times\cdots\times}\mathbf{T}_{P}^{\ast}\mathcal{M}}_{p\,\text{factors}}\times\underbrace{\mathbf{T}_{P}\mathcal{M}\mathcal{\times\cdots\times}\mathbf{T}_{P}\mathcal{M}}_{q\,\text{factors}}\to\mathbb{R}.
\]
A tensor like this can be expanded on a basis

\[
\left\{ \boldsymbol{\theta}^{\mu_{1}}\otimes\cdots\otimes\boldsymbol{\theta}^{\mu_{p}}\otimes\mathbf{e}_{\nu_{1}}\otimes\cdots\otimes\mathbf{e}_{\nu_{q}}\right\} 
\] and its components transform as:
\begin{align}
\overline{W}_{\beta_{1}\cdots\beta_{p}}^{\alpha_{1}\cdots\alpha_{q}}=\left(\frac{\bar{\phi}\left(\bar{x}\right)}{\phi\left(x\right)}\right)^{q-p}\,\frac{\partial\bar{x}^{\alpha_{1}}}{\partial x^{\mu_{1}}}\cdots\frac{\partial\bar{x}^{\alpha_{q}}}{\partial x^{\mu_{q}}}\times\ \ \ \ \,\nonumber \\
\times\frac{\partial x^{\nu_{1}}}{\partial\bar{x}^{\beta_{1}}}\cdots\frac{\partial x^{\nu_{p}}}{\partial\bar{x}^{\beta_{p}}}\,\,W_{\nu_{1}\cdots\nu_{p}}^{\mu_{1}\cdots\mu_{q}}\label{eq:09}
\end{align}

under changes of Lyra frames.

\subsection{The Metric }

The differentiable structure over $\mathcal{M}$ presented so far
is not capable of defining scalar products between tangent vectors,
a fact that makes it impossible to determine lengths and angles. Since
it is desired to use Lyra's geometry as a space-time model, it is
necessary to consider an additional structure on $\mathcal{M}$ so
that the inner product of vectors $\left\{ \mathbf{u},\mathbf{v}\right\} \in\mathbf{T}_{P}\mathcal{M}$
is defined. This can be accomplished by equipping the Lyra manifold
with a $\left(0,2\right)$ metric tensor $\mathbf{g}:\mathbf{T}_{P}\mathcal{M}\mathcal{\times}\mathbf{T}_{P}\mathcal{M}\to\mathbb{R}$,
characterized as a bilinear, symmetric and non-degenerate map. By
introducing a reference frame, its components $g_{\mu\nu}$ will be
obtained through the relation $g_{\mu\nu}=\mathbf{g}\left(\mathbf{e}_{\mu},\mathbf{e}_{\nu}\right)$,
and the dot product between the two above vectors will be $\mathbf{g}\left(\mathbf{v},\mathbf{u}\right)=g_{\mu\nu}u^{\,\mu}v^{\,\nu}$.
The components $g_{\mu\nu}$ form a non-singular matrix, whose inverse
components $g^{\mu\nu}$ transform like the components of a $\left(2,0\right)-$tensor. 

The definition of an inner product naturally leads to a canonical
association between $\mathbf{T}_{P}^{*}\mathcal{M}$ and $\mathbf{T}_{P}\mathcal{M}$.
Take a vector $\mathbf{u}\in\mathbf{T}_{P}\mathcal{M}$. A covector
$\mathbf{T}_{P}\mathcal{M}\to\mathbb{R}$ is defined through $\mathbf{g}\left(\mathbf{u},\mathbf{\cdot}\right)$.
Choose a frame, such that $\mathbf{u}=u^{\mu}\mathbf{e}_{\mu}$. Then,
the covector components $u_{\nu}$ , can be obtained by the referred
application to the base vectors of $\mathbf{T}_{P}\mathcal{M}$: $u_{\nu}=\mathbf{g}\left(\mathbf{u},\mathbf{e}_{\nu}\right)=g_{\mu\nu}u^{\nu}$.

As in Riemannian geometry, the matrix of components $g_{\mu\nu}$
has $p$ positive eigenvalues, $q$ negative eigenvalues and
$r$ null eigenvalues, such that $p+q+r=n$, where $n=\dim\mathcal{M}$.
The proportion of each of these quantities determines the signature
$\left(p,q,r\right)$ for the metric that does not depend on the choice
of frames and is constant throughout $\mathcal{M}$, provided that
the metric is non-degenerate and smooth at all points of the manifold.
In the case of $\left(3,0,0\right)$ and $\left(0,3,0\right)$ the
signature is said to be Riemannian whereas $\left(1,3,0\right)$ or
$\left(3,1,0\right)$ are said to be Lorentzian signatures \cite {hawking1973large}.

The introduction of a metric allows the definition of essential physical
elements in the construction of a gravity theory. The first one is
the concept of length. Let be a curve $\lambda:\mathbb{R}\to\mathcal{M}$,
with tangent vectors $\mathbf{v}=\phi\left(x\right)x^{\mu\prime}\mathbf{e}_{\mu}$.
The curve segment $s$ connecting $a$ and $b$ is defined as:
\begin{equation}
s=\int_{a}^{b}dt\,\mathbf{g}\left(\mathbf{v},\mathbf{v}\right)^{1/2}=\int_{a}^{b}dt\sqrt{\phi^{2}g_{\mu\nu}\frac{dx^{\mu}}{dt}\frac{dx^{\nu}}{dt}},\label{eq:09a}
\end{equation}
where $\left[a,b\right]\subset\mathbb{R}$. A metric geodesic is defined
as the curve through points $a$ and $b$ whose distance $s$ is stationary
under fixed infinitesimal variations at the extremes. By performing
the variations in (\ref{eq:09a}) imposing $\delta s=0$, we obtain
the geodesic equation in Lyra manifold:

\begin{align}
& \frac{d^{2}x^{\mu}}{dt^{2}}+\christoffel{\mu}{\alpha\beta}\frac{dx^{\alpha}}{dt}\frac{dx^{\beta}}{dt}+\nonumber \\
& \,\,\,\,+\frac{1}{\phi}\left(\partial_{\alpha}\phi\delta_{\beta}^{\mu}+\partial_{\beta}\phi\delta_{\alpha}^{\mu}-\partial_{\nu}\phi g^{\mu\nu}g_{\alpha\beta}\right)\frac{dx^{\alpha}}{dt}\frac{dx^{\beta}}{dt}=0\label{eq:s21-1}
\end{align}
where $\begin{Bmatrix}\mu\\
\alpha\beta
\end{Bmatrix}$ are the Christoffel symbols.

In Lyra geometry, the presence of the scale influences the volume
invariant element. Starting from the transformation law for the metric
tensor components - which is a $\left(0,2\right)$ tensor - it can
be showed that the determinant of the metric transforms under a change
of frames as: 
\[
\tilde{g}=\frac{\phi^{2n}}{\bar{\phi}^{2n}}\det\left(\frac{\partial x^{\alpha}}{\partial\bar{x}^{\mu}}\right)^{2}g,
\]
where $g:=\det g_{\mu\nu}$ and $n$ is the dimension of the space. Thus, the volume integration element
will be $d^{n}x\phi^{n}\,\sqrt{\left|g\right|}$ and the volume of
a given region $R\subset\mathcal{M}$ will be :
\[
\text{vol}\left(R\right)=\int_{\chi\left(R\right)}d^{n}x\,\phi^{n}\,\sqrt{\left|g\right|}.
\]

\subsection{Linear Connection}

A linear connection over $\mathcal{M}$ is defined as a bilinear map
$\nabla:\mathbf{T}_{P}\mathcal{M}\mathcal{\times}\mathbf{T}_{P}\mathcal{M}\to\mathbf{T}_{P}\mathcal{M}$
relating two tangent vectors $\mathbf{u}$ e $\mathbf{v}$ to a $\left(1,1\right)-$tensor
$\nabla_{\mathbf{u}}\mathbf{v}$ \cite{hicks1965notes}. Let $\mathbf{\mathbf{u}},\mathbf{v},\mathbf{w}\in\mathbf{T}_{P}\mathcal{M}$
and a smooth function $f\in C^{\infty}$. The connection respects
the following properties:\\

\textbf{(i)}~$\nabla_{\mathbf{\mathbf{u}}}\left(\mathbf{v}+\mathbf{w}\right)=\nabla_{\mathbf{\mathbf{u}}}\mathbf{v}+\nabla_{\mathbf{\mathbf{u}}}\mathbf{w}$~;

\textbf{(ii)}~$\nabla_{\mathbf{u}+\mathbf{v}}\left(\mathbf{w}\right)=\nabla_{\mathbf{u}}\mathbf{w}+\nabla_{\mathbf{v}}\mathbf{w}$~;

\textbf{(iii)}~$\nabla_{f\mathbf{u}}\left(\mathbf{v}\right)=f\nabla_{\mathbf{u}}\mathbf{v}$;
and

\textbf{(iv)}~$\nabla_{\mathbf{u}}\left(f\mathbf{v}\right)=\left(\mathbf{u}f\right)\mathbf{v}+f\nabla_{\mathbf{u}}\mathbf{v}$.

~

Let $\sigma:\mathbb{R}\to\mathcal{M}$ and $\lambda:\mathbb{R}\to\mathcal{M}$
be two smooth curves over $\mathcal{M}$ which intersect at the point
$P=\sigma\left(0\right)=\lambda\left(0\right)\in\mathcal{M}$. 

Through
Eq. (\ref{eq:01}), one can define the tangent vectors $\mathbf{v}_{\left(\sigma\right)}$
to $\sigma$ and $\mathbf{u}_{\left(\lambda\right)}$ to $\lambda$,
where $f\in C^{\infty}:\mathcal{M}\to\mathbb{R}$. Taking a specific
frame $\left(U,\chi,\phi\right)$, such that $P\in U$, the vectors
can be represented as $\mathbf{v}_{\left(\sigma\right)}=v_{\left(\sigma\right)}^{\hphantom{\alpha}\mu}\mathbf{e}_{\mu}$
and $\mathbf{u}_{\left(\lambda\right)}=u_{\left(\lambda\right)}^{\hphantom{\alpha}\mu}\mathbf{e}_{\mu}$,
where their respective components are given by $v_{\left(\sigma\right)}^{\hphantom{\alpha}\mu}=\phi\left(x\right)\left(dx^{\mu}/dt\right)\left(0\right)$
and $u_{\left(\lambda\right)}^{\hphantom{\alpha}\mu}=\phi\left(x\right)\left(dx^{\mu}/dt\right)\left(0\right)$.
The operation $\nabla_{\mathbf{u}_{\left(\lambda\right)}}\mathbf{v}_{\left(\sigma\right)}$
represents the change in the vector field $\mathbf{v}_{\left(\sigma\right)}$
in the direction of $\mathbf{u}_{\left(\lambda\right)}$. Therefore,
the related connection naturally brings with it the concept of parallel
vector fields along a curve. If $\nabla_{\mathbf{u}_{\left(\lambda\right)}}\mathbf{v}_{\left(\sigma\right)}=0$,
it is said that $\mathbf{v}_{\left(\sigma\right)}$ is paralleled
in the direction of $\lambda$.

In this frame, the tensor $\nabla_{\mathbf{u}_{\left(\lambda\right)}}\mathbf{v}_{\left(\sigma\right)}$
will be:
\begin{align*}
\nabla_{\mathbf{u}_{\left(\lambda\right)}}\mathbf{v}_{\left(\sigma\right)} & =u_{\left(\lambda\right)}^{\hphantom{\alpha}\nu}\left(\nabla_{\mathbf{e}_{\nu}}v_{\left(\sigma\right)}^{\mu}\right)\mathbf{e}_{\mu}+u_{\left(\lambda\right)}^{\hphantom{\alpha}\nu}v_{\left(\sigma\right)}^{\mu}\Gamma_{\hphantom{\alpha}\mu\nu}^{\alpha}\mathbf{e}_{\alpha}\,\,\,\,\,,
\end{align*}
where the connection coefficients were defined as:
\begin{equation}
\Gamma_{\hphantom{\alpha}\mu\nu}^{\alpha}\mathbf{e}_{\alpha}:=\nabla_{\mathbf{e}_{\nu}}\mathbf{e}_{\mu}\,.\label{eq:10}
\end{equation}
Using expression (\ref{eq:02}), one finds: 
\begin{equation}
\nabla_{\mathbf{u}_{\left(\lambda\right)}}\mathbf{v}_{\left(\sigma\right)}=u_{\left(\lambda\right)}^{\hphantom{\alpha}\nu}\left(\phi^{-1}\partial_{\nu}v_{\left(\sigma\right)}^{\alpha}+\Gamma_{\hphantom{\alpha}\mu\nu}^{\alpha}v_{\left(\sigma\right)}^{\mu}\right)\mathbf{e}_{\alpha}\,.\label{eq:11}
\end{equation}

Therefore, the change in the components of $\mathbf{v}_{\left(\sigma\right)}$
under parallel transport comes as a result of the differential equation:
\begin{equation}
\phi^{-1}u_{\left(\lambda\right)}^{\hphantom{\alpha}\nu}\partial_{\nu}v_{\left(\sigma\right)}^{\alpha}+\Gamma_{\hphantom{\alpha}\mu\nu}^{\alpha}v_{\left(\sigma\right)}^{\mu}u_{\left(\lambda\right)}^{\hphantom{\alpha}\nu}=0.\label{eq:11a}
\end{equation}

\subsubsection{Covariant Derivative}

Consider the curves $\lambda_{\left(\nu\right)}$ generating the base
vectors $\mathbf{e}_{\nu}$ in a given frame. The covariant derivative
of a vector $\mathbf{v}$ in the direction of $\mathbf{e}_{\nu}$
will be written as:
\begin{equation}
\nabla_{\mathbf{e}_{\nu}}\mathbf{v}=\left(\phi^{-1}\partial_{\nu}v^{\alpha}+\Gamma_{\hphantom{\alpha}\mu\nu}^{\alpha}v^{\mu}\right)\mathbf{e}_{\alpha}.\label{eq:11b}
\end{equation}
Its components
\begin{equation}
\nabla_{\nu}v^{\alpha}=\frac{1}{\phi}\partial_{\nu}v^{\alpha}+\Gamma_{\hphantom{\alpha}\mu\nu}^{\alpha}v^{\mu}\,,\label{eq:11c}
\end{equation}
differ from the Riemannian case, by the presence of the factor $\phi^{-1}$
in the ordinary derivative.

The covariant derivative of a covector $\boldsymbol{\omega}$ will
be written as $\nabla_{\mathbf{e}_{\mu}}\boldsymbol{\omega}=\nabla_{\mu}\omega_{\alpha}\,\boldsymbol{\theta}^{\alpha}$.
By using the fact that the covariant derivative of a scalar $f$ is
$\nabla_{\mathbf{u}}f=\mathbf{u}f$ and that the application of a
$\boldsymbol{\omega}$ vector to a tangent vector returns a scalar,
we find that:
\begin{equation}
\nabla_{\mu}\omega_{\nu}=\frac{1}{\phi}\partial_{\mu}\omega_{\nu}-\Gamma_{\hphantom{\alpha}\nu\mu}^{\lambda}\omega_{\lambda}.\label{eq:11d}
\end{equation}

The concept of covariant derivative can be extended to a general $\left(p,q\right)-$tensor
$\mathbf{T}$. In this case, its covariant derivative will be referred
to by $\nabla_{\mathbf{e}_{\alpha}}\mathbf{T}$ and can be expanded
as:
\begin{align*}
\nabla_{\mathbf{e}_{\alpha}}\mathbf{T}= & \nabla_{\alpha}T_{\mu_{1}\cdots\mu_{p}}^{\nu_{1}\cdots\nu_{q}}\left(\mathbf{e}_{\alpha}\otimes\boldsymbol{\theta}^{\mu_{1}}\otimes\cdots\otimes\boldsymbol{\theta}^{\mu_{p}}\otimes\mathbf{e}_{\nu_{1}}\otimes\cdots\otimes\mathbf{e}_{\nu_{q}}\right)
\end{align*}
where the components $\nabla_{\alpha}T_{\mu_{1}\cdots\mu_{p}}^{\nu_{1}\cdots\nu_{q}}$
are given by:
\begin{align}
\nabla_{\alpha}T_{\mu_{1}\cdots\mu_{p}}^{\nu_{1}\cdots\nu_{q}}= & \frac{1}{\phi}\partial_{\alpha}T_{\mu_{1}\cdots\mu_{p}}^{\nu_{1}\cdots\nu_{q}}+\Gamma_{\hphantom{\mu}\lambda\alpha}^{\nu_{1}}T_{\mu_{1}\cdots\mu_{p}}^{\lambda\cdots\nu_{q}}\nonumber \\
& +\cdots+\Gamma_{\hphantom{\mu}\lambda\alpha}^{\nu_{q}}T_{\mu_{1}\cdots\mu_{p}}^{\nu_{1}\cdots\lambda}-\Gamma_{\hphantom{\mu}\mu_{1}\alpha}^{\lambda}T_{\lambda\cdots\mu_{p}}^{\nu_{1}\cdots\nu_{q}}\nonumber \\
& -\cdots-\Gamma_{\hphantom{\mu}\mu_{p}\alpha}^{\lambda}T_{\mu_{1}\cdots\lambda}^{\nu_{1}\cdots\nu_{q}}\,.\label{eq:03_05b}
\end{align}

\subsubsection{Autoparallel Curves}

Let $\sigma:\mathbb{R}\to\mathcal{M}$ be a smooth curve over $\mathcal{M}$,
with tangent vector $\mathbf{v}_{\left(\sigma\right)}=v_{\left(\sigma\right)}^{\hphantom{\alpha}\mu}\mathbf{e}_{\mu}$.
This will be called an autoparallel curve if $\mathbf{v}_{\left(\sigma\right)}$
is parallel transported in relation to its own generating curve $\sigma$;
that is, $\nabla_{\mathbf{v}_{\left(\sigma\right)}}\mathbf{v}_{\left(\sigma\right)}=0$.
With the Eq. (\ref{eq:11a}) and the expression for the components
of the tangent vector in terms of the coordinates $v_{\left(\sigma\right)}^{\hphantom{\alpha}\mu}=\phi\left(x\right)\left(dx^{\mu}/dt\right)$,
we find the differential equation for a autoparallel curve is:

\begin{equation}
\frac{d^{2}x^{\nu}}{dt^{2}}+\left(\phi\Gamma_{\hphantom{\alpha}\rho\mu}^{\nu}+\nabla_{\mu}\phi\delta_{\rho}^{\nu}\right)\frac{dx^{\mu}}{dt}\frac{dx^{\rho}}{dt}=0,\label{eq:12-1}
\end{equation}
where $\nabla_{\mu}\phi:=\phi^{-1}\partial_{\mu}\phi.$

\subsubsection{Curvature and Torsion}

The curvature tensor is defined by the linear map $\mathbf{R}:\mathbf{T}_{P}\mathcal{M}\mathcal{\times}\mathbf{T}_{P}\mathcal{M}\mathcal{\times}\mathbf{T}_{P}\mathcal{M}\to\mathbf{T}_{P}\mathcal{M}$
through \cite{kobayashi1963foundations}:
\begin{equation}
\mathbf{R}\left(\mathbf{\mathbf{u}},\mathbf{v}\right)\mathbf{w}=\nabla_{\mathbf{u}}\nabla_{\mathbf{v}}\mathbf{w}-\nabla_{\mathbf{v}}\nabla_{\mathbf{u}}\mathbf{w}-\nabla_{\left[\mathbf{v},\mathbf{u}\right]}\mathbf{w}.\label{eq:12}
\end{equation}
It is clear from this equation the antisymmetric property $\mathbf{R}\left(\mathbf{\mathbf{u}},\mathbf{v}\right)\mathbf{w}=-\mathbf{R}\left(\mathbf{\mathbf{v}},\mathbf{u}\right)\mathbf{w}$.
The curvature components can be obtained through the expression:
\begin{align}
R_{\hphantom{\alpha}\mu\nu\alpha}^{\lambda}= & \frac{1}{\phi^{2}}\partial_{\mu}\left(\phi\Gamma_{\hphantom{\alpha}\alpha\nu}^{\lambda}\right)-\frac{1}{\phi^{2}}\partial_{\nu}\left(\phi\Gamma_{\hphantom{\alpha}\alpha\mu}^{\lambda}\right)\nonumber \\
& \,\,\,\,\,\,\,+\Gamma_{\hphantom{\alpha}\alpha\nu}^{\rho}\Gamma_{\hphantom{\alpha}\rho\mu}^{\lambda}-\Gamma_{\hphantom{\alpha}\alpha\mu}^{\rho}\Gamma_{\hphantom{\alpha}\rho\nu}^{\lambda}\label{eq:15}
\end{align}

The torsion tensor is defined as a map: \[\boldsymbol{\tau}:\mathbf{T}_{P}\mathcal{M}\mathcal{\times}\mathbf{T}_{P}\mathcal{M}\to\mathbf{T}_{P}\mathcal{M}\]
expressed as \cite{kobayashi1963foundations}:
\begin{equation}
\boldsymbol{\tau}\left(\mathbf{\mathbf{u}},\mathbf{v}\right)=\nabla_{\mathbf{u}}\mathbf{v}-\nabla_{\mathbf{v}}\mathbf{u}-\left[\mathbf{u},\mathbf{v}\right],\label{eq:13}
\end{equation}
from where it can be directly checked that $\boldsymbol{\tau}\left(\mathbf{\mathbf{u}},\mathbf{v}\right)=-\boldsymbol{\tau}\left(\mathbf{v},\mathbf{\mathbf{u}}\right)$.
The torsion tensor components in a local frame $\tau_{\hphantom{\alpha}\mu\nu}^{\alpha}$
are obtained applying the expression (\ref{eq:13}) to the vector
basis:
\begin{align*}
\tau_{\hphantom{\alpha}\mu\nu}^{\alpha}\mathbf{e}_{\alpha} & =\boldsymbol{\tau}\left(\mathbf{e}_{\mu},\mathbf{e}_{\nu}\right)=\left(\Gamma_{\hphantom{\alpha}\nu\mu}^{\alpha}-\Gamma_{\hphantom{\alpha}\mu\nu}^{\alpha}-\gamma_{\hphantom{\alpha}\mu\nu}^{\alpha}\right)\mathbf{e}_{\alpha},
\end{align*}
where $\gamma_{\hphantom{\alpha}\mu\nu}^{\alpha}$ are the structure
constants of the Lie algebra of local basis vectors, given in (\ref{eq:03b}).
Furthermore, the components of the torsion tensor will be:
\begin{equation}
\tau_{\hphantom{\alpha}\mu\nu}^{\alpha}=\Gamma_{\hphantom{\alpha}\nu\mu}^{\alpha}-\Gamma_{\hphantom{\alpha}\mu\nu}^{\alpha}+\phi^{-1}\left(\nabla_{\mu}\phi\delta_{\nu}^{\alpha}-\nabla_{\nu}\phi\delta_{\mu}^{\alpha}\right).\label{eq:14}
\end{equation}

Starting from Eqs. (\ref{eq:13}) and (\ref{eq:12}), considering
$\mathfrak{S}$ as a cyclic sum symbol, it can be shown that curvature
and torsion respect \cite {kobayashi1963foundations}:
\begin{enumerate}
	\item the first Bianchi identity:
	\begin{equation}
	\mathfrak{S}\left[\mathbf{R}\left(\mathbf{\mathbf{u}},\mathbf{v}\right)\mathbf{w}\right]=\mathfrak{S}\left[\boldsymbol{\tau}\left(\mathbf{\boldsymbol{\tau}\left(\mathbf{\mathbf{u}},\mathbf{v}\right)},\mathbf{w}\right)\right]+\left(\nabla_{\mathbf{u}}\boldsymbol{\tau}\right)\left(\mathbf{v},\mathbf{w}\right);\label{eq:16}
	\end{equation}
	\item the second Bianchi identity:
	\begin{equation}
	\mathfrak{S}\left[\left(\nabla_{\mathbf{u}}\mathbf{R}\right)\left(\mathbf{v},\mathbf{w}\right)\right]+\mathbf{R}\left(\mathbf{\boldsymbol{\tau}\left(\mathbf{\mathbf{u}},\mathbf{v}\right)},\mathbf{w}\right)=0\,.\label{eq:17}
	\end{equation}
\end{enumerate}

\section{Assumptions on Lyra Spacetime\label{sec:Espa=0000E7o-Tempo-de-Lyra}}

The geometric structure of the Lyra manifold $\left(\mathcal{M},\mathbf{g},\phi,\mathbf{\nabla}\right)$
can be used as a model for relativistic space-time. A four-dimensional
manifold is adopted, equipped with a Lorentzian metric of signature
$\left(+,-,-,-\right)$. A line element in this spacetime model will
be:
\[
ds^{2}=g_{\mu\nu}\,\phi dx^{\mu}\phi dx^{\nu},
\]
where $ds^{2}$ is invariant by general coordinate and scale transformations.

Although the geometric structure of Lyra manifold is very rich in
terms of geometric conceptualization, it depends on many degrees of
freedom to describe the properties of space-time. Looking for an analogue
in the case of pseudo-Riemannian manifolds, it can be concluded that,
although the theories of Einstein-Cartan and Weyl, for example, present
more daring proposals for geometrization of physics, it is the Theory
of General Relativity that best provides a description, according
to scientific method, of the gravitational effects on the scales of
the solar system. The spacetime of General Relativity is obtained
from the hypotheses of metric compatibility and torsion-free manifold,
which considerably simplifies the geometric structure and field equations.

Likewise, it is desirable to impose a few hypotheses to restrict the
degrees of freedom of Lyra manifold. We would like that Lyra gravitodynamics
steams from Lyra manifold in the same way General Relativity is the
theory of gravity coming from the geometry in a pseudo-Riemannian
manifold. That is, we require Lyra spacetime to be a metric-compatible 

\begin{equation}
\nabla_{\mathbf{u}}\mathbf{g}\left(\mathbf{v},\mathbf{w}\right)=0\label{eq:18}
\end{equation}
and torsion-free spacetime:
\begin{equation}
\nabla_{\mathbf{u}}\mathbf{v}-\nabla_{\mathbf{v}}\mathbf{u}=\left[\mathbf{u},\mathbf{v}\right].\label{eq:19}
\end{equation}
As a consequence, the connection will be dependent only on the metric
and the scale function. One verify this claim by taking into account
the possible permutations of (\ref{eq:18}) and building a tensor
$\mathfrak{W}\left[\nabla_{\mathbf{u}}\mathbf{g}\left(\mathbf{v},\mathbf{w}\right)\right]$
defined by:
\[
\mathfrak{W}\left[\nabla_{\mathbf{u}}\mathbf{g}\left(\mathbf{v},\mathbf{w}\right)\right]:=\nabla_{\mathbf{u}}\mathbf{g}\left(\mathbf{v},\mathbf{w}\right)+\nabla_{\mathbf{v}}\mathbf{g}\left(\mathbf{u},\mathbf{w}\right)-\nabla_{\mathbf{w}}\mathbf{g}\left(\mathbf{u},\mathbf{v}\right)\,,
\]
which is symmetric with respect to $\mathbf{u}$ and $\mathbf{v}$.
Since the connection respects Leibniz's rule, the covariant directional
derivative applied to the metric can be rewritten as \cite{kobayashi1963foundations}:
\[
\nabla_{\mathbf{u}}\mathbf{g}\left(\mathbf{v},\mathbf{w}\right)=\mathbf{u}\left[\mathbf{g}\left(\mathbf{v},\mathbf{w}\right)\right]-\mathbf{g}\left(\mathbf{\nabla_{\mathbf{u}}v},\mathbf{w}\right)-\mathbf{g}\left(\mathbf{v},\nabla_{\mathbf{u}}\mathbf{w}\right)\,.
\]
This relation together with the metric-compatiblity and torsion-free
conditions, defined on Eqs. (\ref{eq:18}) and (\ref{eq:19}), cast $\mathfrak{W}\left[\nabla_{\mathbf{u}}\mathbf{g}\left(\mathbf{v},\mathbf{w}\right)\right]$
in the form:
\begin{align*}
2\mathbf{g}\left(\nabla_{\mathbf{v}}\mathbf{u},\mathbf{w}\right)= & \mathbf{u}\left[\mathbf{g}\left(\mathbf{v},\mathbf{w}\right)\right]+\mathbf{v}\left[\mathbf{g}\left(\mathbf{w},\mathbf{u}\right)\right]\\
& -\mathbf{w}\left[\mathbf{g}\left(\mathbf{u},\mathbf{v}\right)\right]+\mathbf{g}\left(\left[\mathbf{v},\mathbf{u}\right],\mathbf{w}\right)\\
& +\mathbf{g}\left(\left[\mathbf{w},\mathbf{u}\right],\mathbf{v}\right)+\mathbf{g}\left(\left[\mathbf{w},\mathbf{v}\right],\mathbf{u}\right)\,.
\end{align*}
The components of this equation can be obtained by taking a frame
where $\mathbf{u}=\mathbf{e}_{\mu}$, $\mathbf{v}=\mathbf{e}_{\nu}$
and $\mathbf{w}=\mathbf{e}_{\lambda}$, along with Eqs. (\ref{eq:03b})
and (\ref{eq:10}):
\begin{align}
\Gamma_{\hphantom{\alpha}\mu\nu}^{\alpha}= & \frac{1}{\phi}\christoffel{\alpha}{\mu\nu}+\phi^{-1}\left(\nabla_{\mu}\phi\delta_{\nu}^{\alpha}-g_{\mu\nu}\nabla^{\alpha}\phi\right).\label{eq:20}
\end{align}

Another important property of the connection coefficients is the following.
By replacing them (\ref{eq:20}) in the Eq. (\ref{eq:12-1}), the
autoparallel curve equation is reduced to:

\begin{align}
& \frac{d^{2}x^{\alpha}}{dt^{2}}+\christoffel{\alpha}{\mu\nu}\frac{dx^{\mu}}{dt}\frac{dx^{\nu}}{dt}\nonumber \\
& \,\,+\frac{1}{\phi}\left(\partial_{\mu}\phi\delta_{\nu}^{\alpha}+\partial_{\nu}\phi\delta_{\mu}^{\alpha}-g^{\alpha\lambda}g_{\mu\nu}\partial_{\lambda}\phi\right)\frac{dx^{\mu}}{dt}\frac{dx^{\nu}}{dt}=0\label{eq:21}
\end{align}
which is exactly the Eq. (\ref{eq:s21-1}) of the geodesic curve.
Consequently, \emph{in a metric-compatible and torsion-free Lyra manifold, metric and affine geodesics are equivalent}.

Just like it happens for the connection coefficients, the curvature
tensor components (\ref{eq:15}) in a specific frame depend on both
the metric and the scale function. Plugging Eq. (\ref{eq:20}) into
Eq. (\ref{eq:15}), one finds:

\begin{align}
R_{\hphantom{\mu}\mu\nu\beta}^{\alpha}= & \frac{1}{\phi^{2}}\mathcal{R}_{\hphantom{\mu}\mu\nu\beta}^{\alpha}-\frac{2}{\phi}\delta_{[\nu}^{\alpha}\nabla_{\beta]}\nabla_{\mu}\phi+\frac{2}{\phi}g_{\mu[\nu}\nabla_{\beta]}\nabla^{\alpha}\phi\nonumber \\
& +\frac{2}{\phi^{2}}\delta_{[\nu}^{\alpha}g_{\beta]\mu}\nabla_{\lambda}\phi\nabla^{\lambda}\phi\,\,,\label{eq:22}
\end{align}
where $\nabla_{\mu}\phi:=\phi^{-1}\partial_{\mu}\phi$ and $\mathcal{R}_{\hphantom{\mu}\mu\nu\beta}^{\alpha}$
is the curvature tensor evaluated with the Christoffel symbols of
General Relativity \cite{misner1973gravitation}. Lyra's analogue for
the Ricci tensor is obtained by contracting the indexes $\alpha$
e $\beta$ of (\ref{eq:22}):

\begin{align}
R_{\mu\nu}= & \frac{1}{\phi^{2}}\mathcal{R}_{\mu\nu}+\frac{2}{\phi}\nabla_{\nu}\nabla_{\mu}\phi+\frac{1}{\phi}g_{\mu\nu}\nabla_{\alpha}\nabla^{\alpha}\phi\nonumber \\
& -\frac{3}{\phi^{2}}g_{\mu\nu}\nabla_{\lambda}\phi\nabla^{\lambda}\phi\,.\label{eq:23}
\end{align}
The curvature scalar for Lyra manifold is the trace of $R_{\mu\nu}$:
\begin{equation}
R=\frac{1}{\phi^{2}}\mathcal{R}+\frac{6}{\phi}\nabla_{\lambda}\nabla^{\lambda}\phi-\frac{12}{\phi^{2}}\nabla_{\lambda}\phi\nabla^{\lambda}\phi\,.\label{eq:24}
\end{equation}

Lyra's curvature tensor has the following antisymmetry properties:
\begin{equation}
R_{\mu\nu\alpha\beta}=R_{[\mu\nu]\alpha\beta}\,\,\,\,\,\text{and}\,\,\,\,\,\,R_{\mu\nu\alpha\beta}=R_{\mu\nu[\alpha\beta]}\label{eq:25}
\end{equation}
Moreover, it is symmetrical under the exchange between the first two
indexes and the last two indexes:
\begin{equation}
R_{\mu\nu\alpha\beta}=R_{\alpha\beta\mu\nu}.\label{eq:26}
\end{equation}
Once the torsion-free condition is assumed, the Bianchi identities
(\ref{eq:16}) and (\ref{eq:17}) projected in a specific frame are
expressed as:
\begin{equation}
R_{\hphantom{\mu}\mu\nu\beta}^{\alpha}+R_{\hphantom{\mu}\nu\beta\mu}^{\alpha}+R_{\hphantom{\mu}\beta\mu\nu}^{\alpha}=0\label{eq:27}
\end{equation}
and
\begin{equation}
\nabla_{\lambda}R_{\hphantom{\mu}\mu\nu\beta}^{\alpha}+\nabla_{\nu}R_{\hphantom{\mu}\mu\beta\lambda}^{\alpha}+\nabla_{\beta}R_{\hphantom{\mu}\mu\lambda\nu}^{\alpha}=0\,.\label{eq:28}
\end{equation}

From the contracted form for of Eq. (\ref{eq:28}), one finds that
\begin{equation}
\nabla^{\mu}G_{\mu\nu}=0\label{eq:29}
\end{equation}
where the Einstein's tensor analogue for Lyra manifold $G_{\mu\nu}$
was defined as:
\begin{equation}
G_{\mu\nu}:=R_{\mu\nu}-\frac{1}{2}g_{\mu\nu}R,\label{eq:31}
\end{equation}
and, through Eqs. (\ref{eq:23}) and (\ref{eq:24}), it can be rewritten
as:
\begin{align}
G_{\mu\nu}= & \frac{1}{\phi^{2}}\mathcal{G}_{\mu\nu}+\frac{2}{\phi}\nabla_{\nu}\nabla_{\mu}\phi-\frac{2}{\phi}g_{\mu\nu}\nabla_{\lambda}\nabla^{\lambda}\phi\nonumber \\
& +\frac{3}{\phi^{2}}g_{\mu\nu}\nabla_{\lambda}\phi\nabla^{\lambda}\phi\label{eq:32}
\end{align}
where $\mathcal{G}_{\mu\nu}:=\mathcal{R}_{\mu\nu}-\frac{1}{2}g_{\mu\nu}\mathcal{R}$
is the Einstein tensor in General Relativity.

\section{Lyra Scalar-Tensor Theory of Gravity\label{sec:Equacoes-de-Campo}}

The framework presented so far allows the construction of a scalar-tensor
theory of gravity where the basic fields are the reference frame scalar
function $\phi$ and the manifold metric $g_{\mu\nu}$. The action
from which the field equations emerge via the variational principle
will be taken as direct generalization of Einstein-Hilbert action.
We take as ingredients the Lyra's invariant integration element for
a 4-dimensional space-time of Lorentzian metric $d^{4}x\phi^{4}\,\sqrt{-g}$,
the scalar curvature (\ref{eq:24}) plus a Lagrangian density $\mathcal{L}_{\text{m}}=\mathcal{L}_{\text{m}}\left(\psi_{i},\nabla\psi_{i},\phi,\nabla\phi\right)$,
admittedly a Lyra scalar, which describes the contribution of matter
terms. Here, it is assumed that $\mathcal{L}_{\text{m}}$ also depends
on the scale function, in addition to the $\psi_{i}$ matter fields
and their first derivatives. Thus, the action will be:
\begin{align}
S= & \int d^{4}x\,\phi^{4}\,\sqrt{-g}\left[\mathcal{L}_{\text{m}}\left(\psi_{i},\nabla\psi_{i},\phi,\nabla\phi\right)\right.\nonumber \\
& \left.-\frac{1}{16\pi G}\left(\frac{1}{\phi^{2}}\mathcal{R}+\frac{6}{\phi}\nabla_{\lambda}\nabla^{\lambda}\phi-\frac{12}{\phi^{2}}\nabla_{\lambda}\phi\nabla^{\lambda}\phi\right)\right]\label{eq:33}
\end{align}
The variation of (\ref{eq:33}) with respect to the components of
the metric tensor and the scale function leads respectively to the
equations:

\begin{align}
\mathcal{R}_{\mu\nu}-\frac{1}{2}g_{\mu\nu}\mathcal{R}+2\phi\nabla_{(\mu}\nabla_{\nu)}\phi\,\,\,\,\,\,\,\,\nonumber \\
\,\,\,\,\,-2\phi g_{\mu\nu}\nabla_{\lambda}\nabla^{\lambda}\phi+3g_{\mu\nu}\nabla_{\lambda}\phi\nabla^{\lambda}\phi & =-8\pi G\phi^{2}T_{\mu\nu}\label{eq:34}
\end{align}
and
\begin{equation}
6\phi\nabla_{\mu}\nabla^{\mu}\phi-12\nabla_{\alpha}\phi\nabla^{\alpha}\phi+\mathcal{R}=-8\pi G\phi^{2}M\label{eq:35}
\end{equation}
where the stress-energy tensor $T_{\mu\nu}$ was defined as:
\begin{equation}
T_{\mu\nu}=-2\frac{\delta\mathcal{L}_{m}}{\delta g^{\mu\nu}}+g_{\mu\nu}\mathcal{L}_{m}\label{eq:34b}
\end{equation}
and the Lyra scalar $M$ is:
\begin{equation}
M=-4\mathcal{L}_{\text{m}}-\phi\left(\frac{\partial\mathcal{L}_{\text{m}}}{\partial\phi}-\nabla_{\alpha}\frac{\partial\mathcal{L}_{\text{m}}}{\partial\left(\nabla_{\alpha}\phi\right)}\right)\,.\label{eq:35b}
\end{equation}
In addition, variations with respect to the matter fields $\psi_{i}$
give rise to the covariant form of the Euler-Lagrange equations:
\[
\frac{\partial\mathcal{L}_{\text{m}}}{\partial\psi_{i}}-\nabla_{\mu}\frac{\partial\mathcal{L}_{\text{m}}}{\partial\left(\nabla_{\mu}\psi_{i}\right)}=0.
\]

Through (\ref{eq:32}), the Eq. (\ref{eq:34}), which is the Lyra
analogue of Einstein's Equations, can be simply rewritten as $G_{\mu\nu}=8\pi GT_{\mu\nu}$.
Therefore, LyST gravity can be interpreted as the direct generalization
of the theory of General Relativity to Lyra space-time, where the
quantities $G_{\mu\nu}$ and $T_{\mu\nu}$ are tensors, not only by
general transformations of coordinates, but also by scale transformations.

\subsection{Newtonian Limit\label{subsec:O-Limite-Newtoniano}}

The field equations for the LyST theory of gravity must be consistent
with the gravitational phenomenology accessible to experiments and
observations. As is known, the gravitational effects on celestial
mechanics are, with the exception of Mercury's perihelion precession
and light deflection, perfectly described by Newtonian gravity at
the level of the solar system within observational uncertainties.
A successful gravitational theory must, therefore, recover Newton's
equations on a given scale of validity, coined here as \emph{Newtonian 
Limit}.


Under the phenomenological aspect, three basic requirements define
the Newtonian limit to a gravity theory \cite {carroll1997lecture}.
The first is to keep the focus on non-relativistic movements, in such
a way that the spatial components of the $4-$velocity of the test
particles can be neglected, within the required approximation order.
The second condition requires the gravitational field to be static,
which means that the time derivatives of metric that appear in the
Christoffel symbol, as well as the time derivative of $\phi$, can
be neglected. Thus, the equation for geodesics (\ref{eq:21}) is reduced
to: 
\begin{equation}
\frac{d^{2}x^{\mu}}{dt^{2}}=\frac{1}{2}g^{\alpha\mu}\partial_{\alpha}g_{00}+g^{\mu\nu}\frac{1}{\phi}\partial_{\nu}\phi g_{00}.\label{eq:36}
\end{equation}
Finally, the third requirement establishes that the gravitational
field is weak and, as a consequence, must be interpreted as a perturbation
to Minkowski's space-time. In LyST, this limit is obtained based on
the simultaneous assumption of two conditions:
\begin{equation}
g_{\mu\nu}\approx\eta_{\mu\nu}+h_{\mu\nu}\,\,\,\,\,\,\text{ and }\,\,\,\,\,\,\phi\approx1+\delta\phi\left(x\right)\,,\label{eq:37}
\end{equation}
where $h_{\mu\nu}\ll\eta_{\mu\nu}$ and $\delta\phi\left(x\right)\ll1$.
The contravariant components of $h^{\alpha\beta}$ are obtained by
imposing the reciprocal relation between covariant and contravariant
metric coefficients. Considering Eqs. (\ref{eq:37}) in (\ref{eq:36}),
keeping only the first order terms and focusing on the spatial components,
one finds $\ddot{\mathbf{x}}=-\boldsymbol{\nabla}U$, where $U$ is
the Newtonian potential that depends on both the metric and the $\delta\phi$
field:
\begin{equation}
U\equiv\frac{1}{2}h_{00}+\delta\phi.\label{eq:38}
\end{equation}
Once this important finding has been obtained, attention should be
turned to the field equations. By evaluating the trace of Eq. (\ref{eq:34}),
one can cast it into the form:

\begin{align}
& \mathcal{R}_{\mu\nu}+2\phi\nabla_{\mu}\nabla_{\nu}\phi+g_{\mu\nu}\phi\nabla_{\lambda}\nabla^{\lambda}\phi\,\,\,\,\,\nonumber \\
& \,\,-3g_{\mu\nu}\nabla^{\lambda}\phi\nabla_{\lambda}\phi=8\pi G\phi^{2}\left(T_{\mu\nu}-\frac{1}{2}g_{\mu\nu}T\right)\label{eq:39}
\end{align}
In the case of low velocities and weak fields, the gravitational field
is dominated by the time components of (\ref{eq:39}). The stress-energy
momentum tensor, in this case, is dominated by the energy density
$T\approx T_{00}=\rho$, such that the sector $00$ in (\ref{eq:39})
is reduced to:
\begin{equation}
\mathcal{R}_{00}-\nabla^{2}\left(\delta\phi\right)=-4\pi G\rho.\label{eq:40}
\end{equation}
The expression for $\mathcal{R}_{00}$ can be obtained using the equation
for the Ricci tensor in terms of Christoffel's symbols and applying
the conditions that define the Newtonian limit:
\begin{equation}
\mathcal{R}_{00}\approx-\frac{1}{2}\nabla^{2}h_{00}.\label{eq:41}
\end{equation}
Replacing (\ref{eq:41}) in (\ref{eq:40}), one obtains:
\begin{equation}
\nabla^{2}\left(\frac{1}{2}h_{00}+\delta\phi\right)=4\pi G\rho,\label{eq:ln16}
\end{equation}
which shows that the Newtonian potential obtained in (\ref{eq:38}) respects
the Poisson Equation.

\section{ Vacuum Spherically Symmetric Solutions \label{sec:vac-sph}} 

A first case of interest is the one in which space-time is stationary
and spherically symmetric. In General Relativity, the solutions
of Einstein's equations around a stationary black hole can be written
with a metric of the form:
\begin{equation}
g_{\mu\nu}=diag\left(\alpha\left(r\right),-\alpha^{-1}\left(r\right),-r^{2},-r^{2}\sin^{2}\theta\right)\label{eq:ga10}
\end{equation}
Even in the presence of the cosmological constant in the field
equations of General Relativity, this shape for the line element
remains unchanged. Since the observations points to a very significant correspondence
between gravitational phenomena and GR predictions, it must surely
be assumed that a gravitational theory in the Lyra manifold must recover
Einstein's equations on scales of the observed data. Following this
point of view, a working approach can be formulated for the Lyra manifold,
assuming a metric such as (\ref{eq:ga10}), called \emph{Schwarzschild-type}
solutions. Furthermore, it is considered $\phi=\phi\left(r\right)$,
in order to maintain the scale function the invariant of SO(3). This
is not a more general approach, but this consideration greatly simplifies
the work and, as will be shown later, gives rise to solutions that
tend to the Schwarzschild space within specific limits.

The field equations in this framework are reduced
to: 

\begin{subequations}\label{eq:ga11} 
	\begin{align}
	\frac{\alpha'\,\phi'}{\alpha\,\phi}-\frac{1}{\alpha r^{2}}+\frac{1}{r^{2}}+\frac{\alpha'}{\alpha r}+\frac{4\phi'}{r\phi}+\frac{2\phi''}{\phi}-\frac{\phi'^{2}}{\phi^{2}} & =0\label{eq:ga11a}\\
	-\frac{\alpha'\,\phi'}{\alpha\,\phi}+\frac{1}{\alpha r^{2}}-\frac{1}{r^{2}}-\frac{\alpha'}{\alpha r}-\frac{4\phi'}{r\phi}-\frac{3\phi'^{2}}{\phi^{2}} & =0\label{eq:ga11b}\\
	-\frac{\alpha''}{2\alpha}-2\frac{\alpha'\phi'}{\alpha\phi}-\frac{\alpha'}{\alpha r}-\frac{2\phi'}{r\phi}-\frac{2\phi''}{\phi}+\frac{\phi'^{2}}{\phi^{2}} & =0\label{eq:ga11c}
	\end{align}
\end{subequations} Adding the Eq. (\ref{eq:ga11a}) to (\ref{eq:ga11b}),
one finds: 
\begin{equation}
\frac{\phi''}{\phi}-2\frac{\phi'^{2}}{\phi^{2}}=0.\label{eq:ga12}
\end{equation}

Solving this differential equation for $\phi$, it results: 
\begin{equation}
\phi\left(r\right)=\frac{r_{0}}{r_{L}-r},\label{eq:ga17}
\end{equation}
where $r_{L}\in\mathbb{R}$ and $r_{0}\in\mathbb{R}$ are integration
constants. Parameter $r_{L}$ is a new constant appearing in the context of LyST: it will be later identified as Lyra radius and will be key to characterize the causal structure of Lyra-based vacuum spherically symmetric solutions. With the solution (\ref{eq:ga17}), it is possible to use
one of the Eqs. (\ref{eq:ga11}) to find $\alpha$. Replacing (\ref{eq:ga17})
in (\ref{eq:ga11b}): 
\[
r_{L}\alpha'+\alpha\frac{r_{L}}{r}\frac{\left(1+2r/r_{L}\right)}{\left(1-r/r_{L}\right)}=-\left(1-\frac{r_{L}}{r}\right)\,.
\]
This is a first order linear differential equation, which solution
depends on a new integration constant $r_s$ :

\begin{subequations}\label{eq:ga25} 
	
	\begin{equation}
	\alpha\left(r\right)=\frac{\left(1-r/r_{L}\right)^{2}\left(1-r_{s}/r\right)}{\left(1-r_{s}/r_{L}\right)}.\label{eq:ga22}
	\end{equation}
	(Parameter $r_{s}$ will be later related to the familiar Schwarzschild radius.)
	
	The next step is to work with the Eq. (\ref{eq:ga17})
	for $\phi$. As argued before, the theory must recover General
	Relativity when the scale function is constant. By writing (\ref{eq:ga17})
	as $\phi=1/\left(r_{L}/r_{0}-r/r_{0}\right)$ we see that $r_{L}$
	and $r_{0}$ must go to $\infty$ while keeping the ratio $r_{L}/r_{0}$
	constant in order to satisfy this condition. Since the $r_{0}$ appears only in the equation for $\phi$, its possible to factorize the ratio $r_{0}/r_{L}$
	in the line element equation; this leds to $ds^{2}=\frac{r_{0}^{\,2}}{r_{L}^{\,2}}\tilde{\phi}^{2}g_{\mu\nu}dx^{\mu}dx^{\nu}$,
	where both $\tilde{\phi}:=1/\left(1-r/r_{L}\right)$ and $g_{\mu\nu}$
	do not depend on $r_{0}$. Because of this, it is possible to work with
	Lyra frames where $r_{0}=r_{L}$, in which case the scale function becomes
	\begin{equation}
	\phi=\frac{1}{1-r/r_{L}}\,.\label{eq:ga20}
	\end{equation}
\end{subequations}

In this approach, it is evident that Schwarzschild solution of GR arises by taking the limit $r_{L}\to\infty$ in Eqs. (\ref{eq:ga25}). Up to this point, there are no additional constraints on $r_{s}$ and
$r_{L}$. The only information about this constants is that they are
the roots of $\alpha\left(r\right)$ in $\mathbb{R}$. For
more insight into the physical meaning of these parameters, let's
study the weak field limit of LyST gravity.

\subsection{Corrections to Newtonian Gravity \label{sec:5.1} }

Under the requirement of spherical symmetry, the Eq. (\ref{eq:36})
reduces to $\ddot{r}=-\partial_{r}U$. The spherically symmetric
solution for LyST gravity is entirely determined by the functions $\alpha\left(r\right)$
and $\phi\left(r\right)$, cf. Eqs. (\ref{eq:ga25}). By expanding them up
to the second order in $r$, one finds $\phi=1+\delta\phi$ and $\alpha=1+h_{00}$,
where:
\begin{equation}
\delta\phi\approx\frac{r}{r_{L}}+\frac{r^{2}}{r_{L}^{2}}\label{eq:ln17}
\end{equation}
and
\begin{align}
h_{00} & \approx3\left(\frac{r_{s}}{r_{L}}+\frac{r_{s}^{2}}{r_{L}^{2}}\right)-\left(1+\frac{r_{s}}{r_{L}}\right)\frac{r_{s}}{r}+\nonumber \\
& \,\,\,\,\,\,\,\,\,\,\,\,\,\,\,\,\,\,\,\,\,\,\,\,-2\frac{r}{r_{L}}\left(1+\frac{3}{2}\frac{r_{s}}{r_{L}}\right)+\frac{r^{2}}{r_{L}^{2}}\,.\label{eq:ln18}
\end{align}
With these expressions, the radial sector of the equation of motion will be: 
\begin{equation}
\ddot{r}=-\frac{1}{2}\frac{r_{s}}{\left(1-\frac{r_{s}}{r_{L}}\right)}\frac{1}{r^{2}}+\frac{3}{2}\frac{r_{s}}{r_{L}^{2}}-\frac{3r}{r_{L}^{2}}\label{eq:ln23}
\end{equation}

The coefficient of the term scaling as $r^{-2}$ in Eq. (\ref{eq:ln23}) may be defined as a new geometrical parameter m:

\begin{equation}
m:=\frac{1}{2}\frac{r_{s}}{\left(1-\frac{r_{s}}{r_{L}}\right)}.\label{eq:ln24}
\end{equation}
This quantity can be interpreted as the mass in the Newtonian description of gravity. The mass parameter is related not only to Schwarzschild radius $r_{s}$ (as is the case of General Relativity), but also to Lyra radius $r_{L}$.
According to this interpretation, the first order effect of $r_{L}$ is simply to change the way $m$ relates to $r_s$. Second order effects
include a constant repulsive acceleration $3r_{s}/2r_{L}^{2}$ and an Anti-de Sitter type attractive term $-3r/r_{L}^{2}$.

\subsection{Properties of the Spherically Symmetric Solution~\label{sec:Propriedades-da-solucao}}

As explained in the previous section, the existence of a Newtonian limit allows the recognition of a parameter $m$ that can be interpreted as the mass. Following this line of reasoning, this parameter must be taken as a real and positive
quantity based on phenomenological principles. Thus, one can rewrite the solution for $\alpha$ in Eq. (\ref{eq:ga25}) by replacing the constant
$r_{s}$ in terms of m and $r_{L}$. The set of solutions will be:
\begin{equation}
\phi\left(r\right)=\frac{1}{1-\frac{r}{r_{L}}}\;\text{and}\,\,\,\alpha\left(r\right)=\left(1-\frac{r}{r_{L}}\right)^{2}\left(1-\frac{2m}{r}+\frac{2m}{r_{L}}\right)\,,\label{eq:ln28-SOLUCAO}
\end{equation}
which, according to Eq. (\ref{eq:05}), lead to the line element:
\begin{align}
ds^{2}= & \left(1-\frac{2m}{r}+\frac{2m}{r_{L}}\right)dt^{2}\nonumber \\
& \,\,\,-\left(1-\frac{r}{r_{L}}\right)^{-4}\left(1-\frac{2m}{r}+\frac{2m}{r_{L}}\right)^{-1}dr^{2}\nonumber \\
& \,\,\,-\left(1-\frac{r}{r_{L}}\right)^{-2}r^{2}d\Omega^{2}\,\label{eq:ln28-elemento}
\end{align}

\subsubsection*{Schwarzschild Limit}

At this point, it should be checked whether  Eqs (\ref{eq:ln28-SOLUCAO})
are physically plausible. It is known that General Relativity must
be the emerging theory from LyST when scale function $\phi$ is constant. The limit $r_{L}$ $\to\pm\infty$ in (\ref{eq:ln28-SOLUCAO}),
leads directly to: 
\begin{equation}
\alpha\left(r\right)\simeq1-\frac{2m}{r}\label{eq:ga28}
\end{equation}
which is the conventional result from GR. This shows the existence of a well-defined Schwarzschild limit steaming from LyST spherically symmetric solution. This very fact brings about the possibility of interpreting the parameter $r_{s}=\left(1/2m+1/r_{L}\right)^{-1}$
as Lyra's generalization of the Schwarzschild radius.

\subsubsection*{Long-Distance Regime}

The form of the metric coefficients in the infinitely remote
regions from the source can be obtained by taking the limit $r\to\infty$ in Eq. (\ref{eq:ln28-SOLUCAO}). This process depends almost exclusively on a quadratic term in $r$, namely $\alpha\left(r\right)\simeq\left(1+2m/r_{L}\right)r^{2}/r_{L}^{2}$.
If Lyra's space-time were completely characterized by the metric,
it could be understood as an emerging \textquotedblleft cosmological
constant\textquotedblright : 
\begin{equation}
\Lambda\simeq-\frac{3}{r_{L}^{2}}\left(1+\frac{2m}{r_{L}}\right),\label{eq:ga31}
\end{equation}
induced by the presence of the Lyra scale.  Moreover, the observed small absolute value of $\Lambda$ would be explained by a large value $r_{L}$, something that is reassuringly consistent with the Schwarzschild limit discussed above.  It is clear from Eq. (\ref{eq:ga31}) that the
sign of $r_{L}$ determines whether the constant $\Lambda$ will be
positive or negative. In the case where $1+2m/r_{L}<0$ the solution
for the metric would be asymptotically de-Sitter; on the other hand, if $1+2m/r_{L}>0$, it would be Anti-de Sitter type solution. 

This is all too compelling. However, one should recall that in LyST gravitational theory, space-time is specified not
only by the metric tensor, as in the case of General Relativity, but also
by the scale function. Considering the line element (\ref{eq:ln28-elemento}),
it turns out that, in fact, Lyra's spherically symmetric space-time
does not recover the characteristics of either de-Sitter or Anti-de-Sitter space-times in the limits of large distances from the source.

\subsection{Classes of Solutions \label{sec:5.3}}

It is considered that $r\in\mathbb{R}_{+}$, the line element (\ref{eq:ln28-elemento})
is singular at the points $r=r_{s}=\left(1/2m+1/r_{L}\right)^{-1}$
and $r=r_{L}$. These values can be both negative or positive. Accordingly, Lyra's spherically symmetric space-time (\ref{eq:ln28-elemento})
may have different properties in different regions of the parameters' range of values. These properties will be explored later on.

The Newtonian limit offers the interpretation of parameter $m$ as the mass of the source; therefore, it must be a real and positive quantity.  Consequently, it should be $r_{s}>0$ for $r_{L}>0$ according to $r_{s}=\left(1/2m+1/r_{L}\right)^{-1}$.

In the range $r_{L}<-2m$, the Schwarzschild radius remains
positive and the region $0<r<\infty$ contains one singular point.

The interval $-2m<r_{L}<0$ does not induce any
singularities in $\left\{ r\in\mathbb{R}|r>0\right\} $. However, the  coefficients of $dt^{2}$
and $dr^{2}$ in the line element change their signs in this range. This type
of mathematical solution has no physical interest and will not be
considered in the rest of the paper.

Assuming that neither $r_{s}$ nor $r_{L}$ diverge, \footnote{For if $r_{L}\to\infty$, the Schwarzschild solution of General Relativity
	is recovered, which is also a solution of LyST field equations, for
	the constant scale. }, 
LyST spherically symmetric solutions can be classified as:
\begin{itemize}
	\item \textbf{Class 1:} $r_{s}>0$ and $r_{L}>0$, space-time with two singularities;
	\item \textbf{Class 2}: $r_{s}>0$ ans $r_{L}<-2m$, space-time with one
	singularity.
\end{itemize}
It is necessary to characterize the divergent points in the line element to learn if they are essential singularities of space-time, or if they are removable divergences arising from particular choices of Lyra frames. This will be done in the upcoming section.

\subsection{Singularities and Causal Structure \label{sec:Singularidades-e-Estrutura}}

LyST Class 1 solution exhibits two singularities, which
seems to indicate a line element with multiple horizons. Thus, this section aims to study these concepts to determine whether there are physical
singularities and, if so, what their nature is. 

A first case of interest is the determination of light-cone distortions, a task that is accomplished through the study of null geodesics. A local light cone is defined by the vanishing of the line element in Eq. (\ref{eq:05}); this corresponds to a geodesics of non-massive particles.
The relation between coordinates $t$ and $r$ for purely radial motion is given by:
\begin{equation}
\frac{dt}{dr}=\pm\left(1-\frac{r}{r_{L}}\right)^{-2}\left(1-\frac{2m}{r}+\frac{2m}{r_{L}}\right)^{-1}.\label{eq:sec01}
\end{equation}
The solution to Eq. (\ref{eq:sec01}) is:
\begin{equation}
t_{\pm}\left(r\right)=c_{1}\pm\left[\frac{r}{\left(1-\frac{r}{r_{L}}\right)}+2m\ln\left(\frac{r}{1-\frac{r}{r_{L}}}-2m\right)\right]\,.\label{eq:sec02}
\end{equation}

The curves with signal $+$ ($-$) are called the \emph{outgoing (ingoing) null geodesics}.
Note that the limit $r_{L}\to\infty$ naturally leads to the result
of General Relativity \cite{dinverno}. LyST Class 1 solution does not tend to Minkowski space-time for regions very distant from the source;  in fact, for $r\gg r_{L}$ and $r\gg r_{s}$,  it approaches an Anti-de-Sitter space-time. This feature directly influences the radial propagation of light beams, since $dr/dt\neq1$ away from the source.

\begin{figure}[t]
	\includegraphics{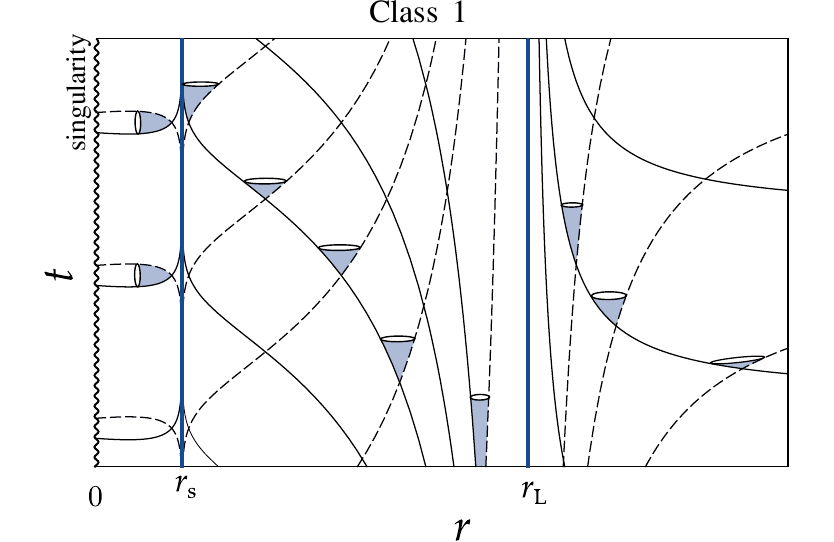}\caption{Null geodesics for the spherically symmetric solution of Class 1 in LyST gravity, in Schwarzschild coordinates. \label{fig:Lyra-schwarzs-diagram-c1}}
\end{figure}

\begin{figure}[t]
	\centering{}\includegraphics{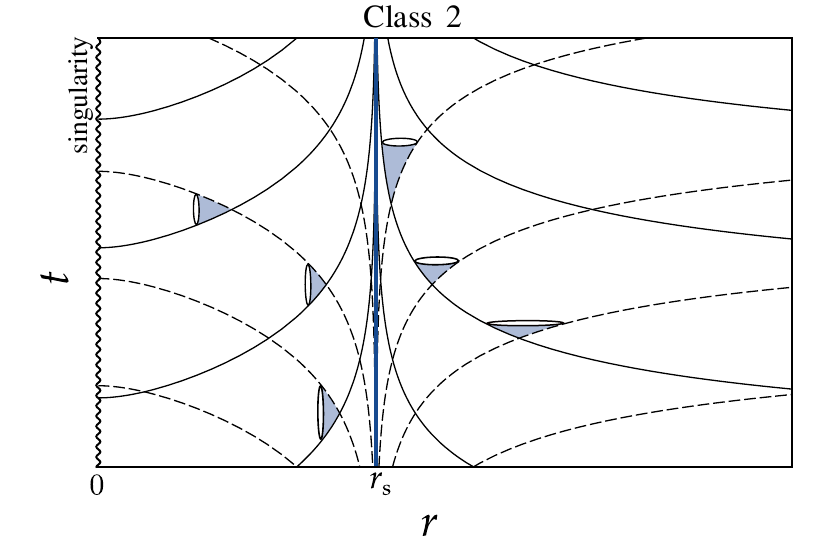}\caption{Space-time diagram for null geodesics in the spherically symmetric solution of Class 2 in LyST gravity with Schwarzschild coordinates.\label{fig:Lyra-schwarzs-diagram-c2}}
\end{figure}

LyST Class 1 space-time diagrams are built from Eq. (\ref{eq:sec02}) through the overlap of ingoing and outgoing null geodesics. This procedure unveils local light cones in strategic regions. Fig. \ref{fig:Lyra-schwarzs-diagram-c1} displays some null-geodesic curves from Eq. (\ref{eq:sec02}) and local light cones in multiple space-time events for the case where $r_{s}<r_{L}$ in LyST Class 1 solutions. We can see a significant distortion of the light cones in the immediate vicinity of both $r_{s}$ and $r_{L}$. This is a direct consequence of the fact that $dt/dr$ diverges at $r_{s}$ and $r_{L}$, the roots of $\alpha\left(r\right)$.The space-time diagram for LyST Class 2 solutions is shown in Fig. \ref{fig:Lyra-schwarzs-diagram-c2}.
It seems that this class give rise to a space-time diagram
qualitatively indistinguishable from that derived from Schwarzschild
solution of General Relativity. 

Through the diagrams in Figs. \ref{fig:Lyra-schwarzs-diagram-c1} and \ref{fig:Lyra-schwarzs-diagram-c2}, it can be checked that the light cones rotate at $90$ degrees counterclockwise to the left of $r_{s}$; in the region $r<r_s$ coordinates $r$ and $t$ switch roles.

The distortion of the light cones in the vicinity of $r=r_{s}$ and $r=r_{L}$ suggests that it takes an infinite
coordinated time $t$ for a test particle to reach the surfaces corresponding to these radii. However, this does not automatically mean that the horizons at $r_{s}$ and $r_{L}$ are essential singularities of LyST spherically symmetric space-time. They might as well be removable coordinate singularities (recall that Kruskal-Szekeres coordinates \cite{hawking1973large} eliminate the divergence in GR's Schwazschild line element at $r=r_{s}$). In order to check this, it is necessary to look for divergences in the gravitational field invariants at the radial values $r_{s}$ and $r_{L}$.

The simplest invariant that can be thought of would be the curvature
scalar $R$. However, in the case of vacuum solutions, the trace of
Einstein's equations shows that $R=0$ in the entire space; therefore, $R$ is useless to assess the nature of the horizons at $r_{s}$ and $r_{L}$. Alternatively,
Kretschmann scalar $K=R_{\mu\nu\alpha\beta}R^{\mu\nu\alpha\beta}$
can be used for this purpose \cite{Ioannis2014Kretshmann}. Utilizing Eqs. (\ref{eq:09}), it follows that:
\begin{equation}
K=\frac{12r_{s}^{2}}{r^{6}}\frac{\left(1-\frac{r}{r_{L}}\right)^{6}}{\left(1-\frac{r_{s}}{r_{L}}\right)^{2}}.\label{eq:sec03}
\end{equation}
Notice that this Kretschmann scalar tends dutifully to the value $K=48m^{2}/r^{6}$ of the ordinary GR's Schwarzschild solution in the limit $r_{L}\rightarrow\pm\infty$. The Kretschmann scalar (\ref{eq:sec03}) remains finite at $r=r_{s}$, which leads to the conclusion that the horizon at $r_{s}$ indicated by Eq. (\ref{eq:sec01}) is not an essential singularity. Consequently, it can be removed through an appropriate choice of Lyra reference frames. On the other hand, Kretschmann invariant for LyST Schwarzschild-type metric approaches zero as $r\rightarrow r_{L}$. Let us investigate the meaning of this fact by considering what happens to a particle in radially free-falling motion.

The equations for the geodesic motion (\ref{eq:sec01}) in a spherically
symmetric Lyra space-time are given by: \begin{subequations}\label{eq:geo19}
	\begin{align}
	& \phi^{2}\alpha\dot{t}=k\,;\label{eq:geo19a}\\
	& \ddot{r}+\alpha^{2}\dot{t}^{2}\left(\frac{\phi'}{\phi}+\frac{\alpha'}{2\alpha}\right)+\nonumber \\
	& \,\,\,\,\,+\dot{r}^{2}\left(\frac{\phi'}{\phi}-\frac{\alpha'}{2\alpha}\right)-\alpha r^{2}\dot{\varphi}^{2}\left(\frac{\phi'}{\phi}+\frac{1}{r}\right)=0\,;\label{eq:geo19b}\\
	& \phi^{2}r^{2}\dot{\varphi}=h\,;\label{eq:geo19c}
	\end{align}
	where $\dot{\left(\;\right)}:=\frac{d}{d\lambda}$ e $\left(\quad\right)^{\prime}:=\frac{d}{dr}$,
	$k$ and $h$ are respectively the specific energy and specific angular
	momentum of the test particle, respectively; and, the functions $\alpha$ and $\phi$
	are given in (\ref{eq:ln28-SOLUCAO}). The components
	$u^{\mu}$ of the four-velocity respect the condition $u_{\mu}u^{\mu}=1$ in natural units. By combining this constraint and Eqs. (\ref{eq:geo19a}-\ref{eq:geo19c}),
	we obtain a differential equation for the evolution of the radial
	coordinate: 
	\begin{equation}
	\frac{k^{2}}{\phi^{2}\alpha}-\frac{\phi^{2}}{\alpha}\dot{r}^{2}-\frac{h^{2}}{\phi^{2}r^{2}}=1\,.\label{eq:geo19d}
	\end{equation}
\end{subequations} In a pure radial motion, the angular momentum
vanishes, and the equation of motion reduces to: 

\noindent 
\begin{equation}
\dot{r}\left(\tau\right)=\pm\left(1-\frac{r}{r_{L}}\right)^{2}\sqrt{\frac{2m}{r}-\frac{2m}{r_{0}}}.\label{eq:sec04}
\end{equation}
Here $r_{0}$ is the value of the radial coordinate where the test particle is at rest. 

\subsection*{Class 1 Solutions}

The allowed types of motion are those that keep the argument of the square root
in Eq. (\ref{eq:sec04}) positive. Thus, the region accessible to time-like
trajectories are those where $r\leq r_{0}$. In order to study the
neighboring regions of $r_{L}$ for $r_{s}<r_{0}\leq r_{L}$, we
take $r_{0}\to r_{L}$ and consider a free-fall trajectory of
a particle that leaves $r_{L}$ from rest. In this case, Eq. (\ref{eq:sec04})
can reads $\dot{r}\left(\tau\right)=-\sqrt{2m}r^{2}\left(1/r-1/r_{L}\right)^{5/2}$
and the proper time for the particle to come out of rest at $r_{L}$
to reach any point of radial coordinate $r_{\text{end}}<r_{L}$ is:
\begin{align*}
\Delta\tau & =-\int_{r_{L}}^{r_{\text{end}}}dr\left(2m\right)^{-1/2}r^{-2}\left(\frac{1}{r}-\frac{1}{r_{L}}\right)^{-5/2}\\
& =\left.\sqrt{\frac{2}{9m}}\left(\frac{1}{r}-\frac{1}{r_{L}}\right)^{-3/2}\right|_{r_{\text{end}}}^{r_{L}}\to \infty\,\,\,.
\end{align*}
In the instance where $r_{0}>r_{L}$, the physically allowed region is
$r_{L}\leq r\leq r_{0}$. The proper time elapsed during a free fall from rest at $r_{0}$ will be — see Eq. (\ref{eq:sec04}):
\[
\Delta\tau=\int_{r_{L}}^{r_{0}}\frac{dr}{\left(1-\frac{r}{r_{L}}\right)^{2}\sqrt{\frac{2m}{r}-\frac{2m}{r_{0}}}}\to\infty\,\,;
\]
that is, a particle starting from rest off the radius $r_{0}>r_{L}$
takes an infinite time to reach $r_{L}$. Along with the fact that
a particle at rest in $r_{L}$ cannot move to values of $r<r_{L}$,
these findings lead to the conclusion that the regions $r<r_{L}$
and $r>r_{L}$ are not in causal contact. 

\subsection*{Class 2 Solutions}

The solutions in this class are constrained to $r_{s}>0 and r_{L}<-2m$. Thus, parameter $r_{L}$ is negative
and cannot be interpreted as a radial distance since it lays outside the validity domain
of the radial coordinate. The singularity at $r_{s}=\left(1/2m+1/r_{L}\right)^{-1}$
can be removed with an appropriate choice of frame as indicated
by the Kretschmann scalar analysis, and the causal structure of this class of solution is equivalent to the already known case of GR's Schwarzschid solution.
The influence of $r_{L}$ with respect to Schwarzschild's event horizon
is to induce changes in the expression involving $r_{s}$ and $m$.

\section{Final Comments  \label{sec:final} }

LyST gravity is a scalar-tensor theory whose fundamental fields are
the metric $g_{\mu\nu}$ and the scale function $\phi$. Unlike established
scalar-tensor theories, LyST is built on a non-Riemannian Lyra manifold.
In fact, the main change on the geometric framework is regarding the definition
of the basis of the tangent and cotangent spaces. It makes the reference
frames dependent on the choice of both the coordinates and the scale
function. In Lyra manifolds, the transformation law of tensors is
distinct from that on the Riemannian manifold.

LyST theory of gravity was built under the geometric hypotheses of
metric-compatibility and torsion-free connection. For this reason,
LyST gravity can be interpreted as the equivalent of General Relativity
in a Lyra manifold. Its field equations were obtained via a Lyra invariant
action (i.e. through an action invariant both under scale and coordinates
transformations) given by the simple generalization of the Einsten-Hilbert
action. By writing the contributions of $\phi$ in the action (\ref{eq:33})
explicitly, we note that LyST is quite different from conventional
theories like Brans-Dicke and its generalizations \cite{sotiriou2012black}.
In the first place, this is true due to the presence of the higher
order derivative term. Furthermore, there is also the factor $\phi^{4}$
that appears to guarantee the symmetry properties of the integration
element, which induces terms dependent on $\phi$ in the field equations
during the process of integration by parts. 

It was found that the linearized and static approximations of LyST
theory tends to a well-defined Newtonian limit for non-relativistic
motion. In fact, we have obtained $U=(1/2)h_{00}+\delta\phi$,
where $\delta\phi$ is the perturbation on the Lyra scale. Trajectories
and field equations were characterized by
the single SO(3) scalar function $U$, 
which is recognized as the Newtonian
gravitational potential, cf. $\ddot{\mathbf{x}} = - \mathbf{\nabla}U $ and Eq. (\ref{eq:ln16}). A remarkable feature of this potential is
its dependence on both the linearized metric 
$h_{00}$ and the scale function perturbation $\delta\phi$,
which is a natural consequence of the fact that the curvature depends on these
two quantities.

LyST gravity presents a spherically symmetric solution
dependent on two parameters $m$ and $r_{L}$. Parameter $m$ is the geometrical mass and $r_{L}$ is a distance related quantity dubbed Lyra radius. LyST spherical metric tends naturally to Schwarzschild solution of General Relativity if $r_{L}\to\pm\infty$, as it can be seen from Eq. (\ref{eq:ln28-SOLUCAO}).
The equations of motion show a well-defined
Newtonian limit: the non-relativistic gravitational regime
is equivalent to Newtonian gravity up to orders $0$ and $1$
of $r_{L}^{\,-1}$. First order effects change the costumary relation between mass and Schwarzschild radius to $r_{s}=\left(1/2m+1/r_{L}\right)^{-1}$,
while higher order effects steaming from terms scalling as  $r_{L}^{\,-2}$  induce de-Sitter-type correction terms to the line element.

LyST spherically symmetric solution splits into two mains cases of physical interests. In Class 1, $r_{L}$ is a positive parameter greater than $r_{s}$, and LyST space-time exhibits two singular points. The singularity at  $r=r_{s}$ is non-essential — just like it happens in General Relativity — and can be removed through an
appropriate choice of frame. On the other hand, there is non-removable horizon at $r_{L}$, where the space-time is divided into two regions without mutual causal contact. LyST spherical metric of Class 2 features $r_{L}<-2m$ so that $r_{L}$ lays outside the domain of radial coordinates. In this case, LyST space-time presents a single Schwarzschild-type singularity and the
only influence of the Lyra scale $\phi$ is to change the expression for the event horizon radius $r_{s}$. 

The fact that LyST gravity exhibits spherically symmetric solutions with a well-defined limit to Schwarzschild solution of General Relativity is a very encouraging finding. Since a significant part of the gravitational phenomena can be described via General Relativity, the uncertainties in the measurements in solar-system-size scales could impose lower limits to the absolute value of LyST parameter $r_{L}$. Future perspectives toward this goal are to build the equations for PPN approximation in the context of LyST. 

On the one hand, local mesuments of gravitational physics constrain LyST free paramenters. On the other hand, large-scale gravitational observations bring an opportunity to the new theoretical framework openned up by LyST. Indeed, GR faces dificulties in cosmology while, for instance, describing the current cosmic acceleration and the $\sigma_{8}$-tension \cite{battye2015tension}. These open problems could be addressed via LyST analogous to FLRW metric plus an SO(3)-invariant scale function $\phi$.

In addition, studies on geodesic motion in spherically symmetric LyST space-time and on gravitational waves in LyST gravity are being carried out. The relaxation of imposing
a torsion-free connection and the metric compatibility condition is
also being investigated in order to determine the Lyra equivalents to Einstein-Cartan
\cite{cartan1923varietes,cartan1924varietes,cartan1925varietes} and
Teleparallel \cite{einstein1928riemann} theories of Gravity.

\begin{acknowledgments}
RCC is greateful to IFT-Unesp for hospitality. EMM thanks CAPES-Brazil
for full financial support (grant 88882.330781/ 2019-01). BMP acknowledges CNPq (Brazil) for partial financial support. The authors thank the referee whose comments helped to improve the paper.
\end{acknowledgments}


\end{document}